\documentclass[3p,authoryear]{elsarticle}
\usepackage{amsmath}
\usepackage{amsthm}
\usepackage{comment}
\usepackage{url}
\usepackage{txfonts}
\usepackage{graphicx}  
\usepackage[utf8]{inputenc}
\usepackage[capitalise]{cleveref}
\usepackage{xcolor}
\usepackage{lscape}
\usepackage{color}
\usepackage{soul}
\usepackage{rotating}
\usepackage{appendix}
\usepackage{subcaption}

\usepackage{multirow}
\usepackage{here}

\usepackage{threeparttable}

\usepackage{bigstrut}
\usepackage[normalem]{ulem}

\newcommand{\add}[1]{\textcolor{black}{#1}}

\begin{document}
\begin{frontmatter}

\title{
Disruption and recovery of the US domestic airline networks during the COVID-19 pandemic}

\address[TSE]{Department of Transdisciplinary Science and Engineering, Tokyo Institute of Technology, Tokyo 152-8550, Japan}

\author[TSE]{Kashin Sugishita\corref{cor}}\ead{sugishita.k.aa@m.titech.ac.jp}\cortext[cor]{Corresponding author}
\author[TSE]{Hiroki Mizutani}
\author[TSE]{Shinya Hanaoka}

\begin{abstract}\quad 
The COVID-19 pandemic has had serious impacts on the airline industry. 
Ensuring that aviation policies in emergent situations both guarantee network connectivity and maintain competition among airlines is crucial in these circumstances. 
To this end, we aimed to understand the network dynamics of individual airlines. 
In this study, we quantitatively reveal the day-to-day dynamics of these US domestic airline networks, comprising 17 airlines, from January 2019 to December 2021. 
Specifically, we applied a framework for analyzing temporal networks, in which the network structure changes over time. 
We found that, first, even though the number of nodes and edges returned to pre-pandemic levels around July 2021, the structure of the entire US domestic airline network remained altered. We also found that the network dynamics varied significantly from airline to airline. Full-service carriers were less flexible in changing their network structure and suffered higher revenue losses. On the contrary, most regional carriers completely shifted to a new structure, which may have contributed to reducing their revenue losses. Low-cost carriers were characterized by more pronounced differences between airlines and drastically changed their network structure immediately after the declaration of a national emergency.  
Finally, we also examined the recovery process and found that the flights connecting airports that are more central and share more common neighbors, and those connecting airports with larger numbers of connections tend to recover earlier for most of the airlines.
\end{abstract} 


\end{frontmatter}

\section{Introduction}
The novel coronavirus, COVID-19, was first identified in China in December 2019 and subsequently caused a global pandemic. 
This pandemic has had a huge impact on many aspects of life, such as the environment, health, and the economy \citep{sarkodie2021global}. 
Regarding the latter, according to the International Monetary Fund (IMF), the cumulative economic losses associated with the pandemic are expected to reach \$13.8 trillion in 2024 \citep{IMF2022economic}.

Among others, the airline industry has been particularly severely affected \citep{suau2020early, sobieralski2020covid, zhang2021impact}. 
The sudden decrease in air travel demand has been much worse than that seen after the 9/11 terrorist attacks and the 2008 financial crisis combined \citep{curley2020coronavirus}. 
Due to this unprecedented decrease in demand, many companies ceased almost all their operations \citep{sun2020did}, leading to many bankruptcies of airlines worldwide \citep{salman2020can}.
Thus far, numerous papers have been published on air transport during the COVID-19 pandemic \citep{sun2021degree, sun2021covid, sun2022covid}. As we will see in Sec.~\ref{sec:review}, some studies have analyzed airline networks, particularly at national and global levels. In these studies, airline networks are typically constructed by aggregating networks of individual airlines. However, each airline has its own corporate strategy, and the dynamics of their networks may vary widely depending on the airline. These kinds of dynamics at the airline level are largely unknown. However, understanding airline-level network dynamics is essential for ensuring aviation policies in emergent situations in order to both guarantee network connectivity and maintain competition among airlines.

\add{In} the present study, we aim to quantitatively reveal the daily dynamics of airline networks during the COVID-19 pandemic for individual airlines. 
Specifically, we analyze the US domestic airline network \add{because it is one of the largest domestic aviation markets in the world. We also took into account the following three points}.
First, reliable and up-to-date daily flight data are available for the US domestic airline network. These data are publicly available from the US Bureau of Transportation Statistics and are based on monthly reports from airlines and are highly reliable \citep{BTS2020air}. We analyzed data from January 2019 to December 2021 to track changes in the network structure before and after the pandemic. Second, there are multiple airlines based in the US. 
In fact, the dataset contains flight data from 17 airlines, including full-service carriers (FSCs), low-cost carriers (LCCs), and regional carriers (RCs). 
\add{Note that these 17 airlines are all the airlines included in the data set.}
This is important for our study purpose, which is to determine the structural network changes for individual airlines. Third, the US is one of the countries most heavily affected by the COVID-19 pandemic. As of May 14, 2022, the cumulative number of COVID-19 cases in the US reached 82 million, which is the highest number of cases worldwide \citep{WHO2022covid}. As such, we decided to analyze the US domestic airline network.

\begin{figure}[bth]
\centering
\includegraphics[width=\linewidth]{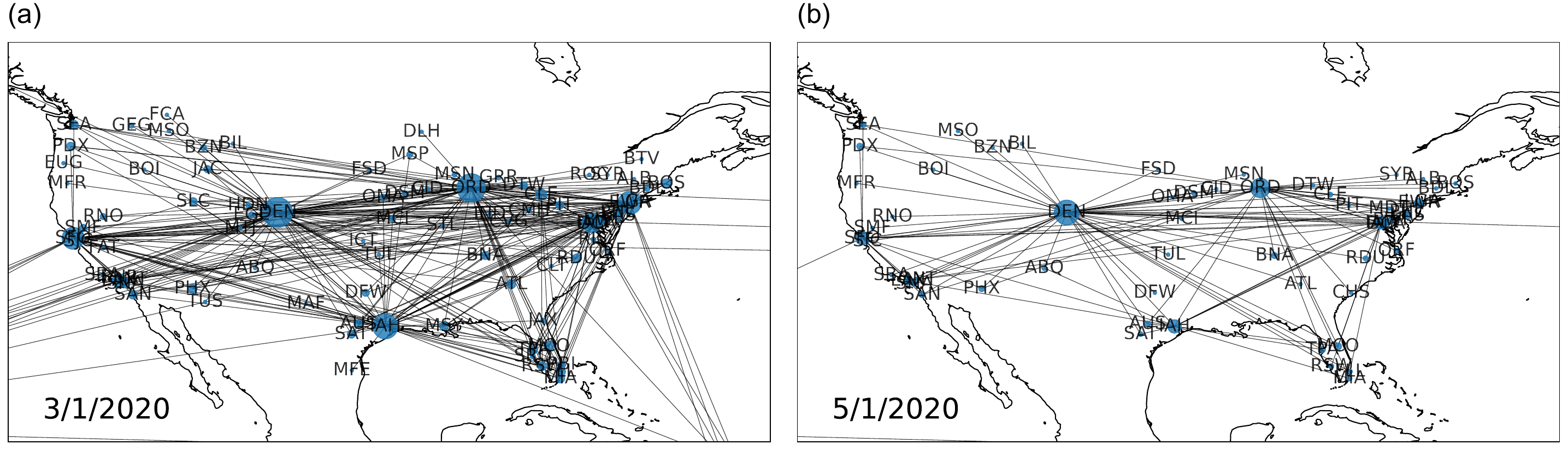}
\caption{\add{Networks of United Airlines on (a) March 1 and (b) May 1, 2020.}}
\label{fig:UA_example}
\end{figure}

In terms of methodology, in addition to some fundamental measures used in network science, such as average degree, average clustering coefficient, and assortativity, we apply a method for analyzing temporal networks, in which the network structure changes over time \citep[e.g.,][]{holme2012temporal, holme2019temporal, masuda2020guidance}. Since airline networks change their structure over time, it is reasonable to consider them temporal networks \citep[e.g.,][]{pan2011path, rocha2017dynamics, sugishita2021recurrence}. In emergent situations such as the COVID-19 pandemic, the network structure could be particularly variable on a daily basis. As an example, we show the networks for United Airlines on March 1 and May 1, 2020, in Fig.~\ref{fig:UA_example}; as seen in the figure, the network shrunk substantially following the pandemic onset. Thus, to quantitatively show the changes in network structure over time, temporal network analyses are required. 
Specifically, in the present study, we analyze the dynamics of airline networks from the perspective of recurrence. Recurrence refers to the return of the network structure to something close to the past structure \citep{masuda2019detecting}. Recurrence in the evolution of network structure has been investigated for some real-world network systems \citep{masuda2019detecting, cruickshank2020characterizing, gelardi2021temporal, sugishita2021recurrence}. However, to the best of our knowledge, it has not been applied to analyze airline networks during the COVID-19 pandemic. This analysis enables us to quantitatively determine how and when the network structure returned to the pre-pandemic structure or if it completely shifted to a new structure.

Our paper proceeds as follows. In Sec.~\ref{sec:review}, we review the relevant literature and further clarify our contributions in this study. In Sec.~\ref{sec:methods}, we describe the methodology used to analyze airline networks. In Sec.~\ref{sec:results},
we show and discuss the results for both the overall US domestic airline network and the networks of individual airlines. In Sec.~\ref{sec:conclusions}, we conclude our work and describe future tasks. In \ref{sec:each_airline_measurements}, we show some additional results of network structure for individual airlines.

\section{Literature review}
\label{sec:review}

To further highlight the contributions in this study, we review the relevant literature on air transport networks during the COVID-19 pandemic.
Even before the COVID-19 pandemic, it was well known that airline networks can be major contributors to the spread of infectious diseases \citep[e.g.,][]{tatem2006global, khan2009spread, pigott2014mapping}.
Now, in the face of a global pandemic, many studies have investigated  \add{air transport networks during the COVID-19 pandemic}.
\add{Several studies analyzed air transport networks during the pandemic from the perspective of network strucutre.
For example, \cite{sun2020did} and \cite{sun2021impact} investigated the impact of the pandemic on air transport networks for the global network and domestic networks of some representative countries.
\cite{li2021impact} analyzed the impact of entry restriction policies on the global air transport network.
\cite{bao2021impact} examined the worldwide air transport network during the pandemic and concluded that the disruption can be considered as a mixture of targeted attacks and random failures on the networks.
\cite{bauranov2021quantifying} also showed that closed airports were small airports on the network's periphery in the case of the US.
\cite{li2021spatiotemporal} showed that the recovery speed of the global network is slower than that of domestic networks because international connectivity is geographically dependent on different policies of travel restrictions among countries.}
Predicting cases and detecting early warning signals of outbreaks from airline networks has \add{also} been a focus. 
For example, \cite{christidis2020predictive} predicted which countries had a high risk of infection using flight data in China. 
Similarly, \cite{daon2020estimating} presented a method to estimate the risk of COVID-19 outbreaks emerging from airline networks on a global scale. \cite{tuite2020estimation} quantified the size of the outbreak in Iran based on flight data between Iran and other countries, and predicted where outbreaks originating in Iran would spread next.
Some studies have also investigated the relationships between COVID-19 cases and measures of network structure such as network density \citep{chu2020detecting}, shortest path length \citep{kuo2021airline}, centrality of nodes \citep{coelho2020global, ribeiro2020severe}, and weight of edges (i.e., number of flights or passengers on each edge) \citep{zhang2020measuring, lau2020association}. 
Additionally, several studies have proposed various models to investigate the virus' spread in airline networks, such as the SIR metapopulation model \citep{hossain2020effects}, SEIR metapopulation model \citep{peirlinck2020outbreak}, and epidemiological models \citep{nikolaou2020identification, riquelme2021contagion}.
\add{From these studies, it can be said that important insights have been gained regarding the structural changes in air transport network at the global, regional, or country levels and their relationship with COVID-19 cases.}

\add{At the airline levels, previous studies have mainly focused on financial conditions during the pandemic.}
In general, airlines can cover only a few months of revenue loss due to high capital costs
\citep{zhang2020exploring}. In fact, some airlines filed for bankruptcy during the pandemic, including Flybe, VA Australia, and Level Europe, among many others \citep{salman2020can, gossling2020pandemics, florido2021effects}.
In emergent situations, governments generally are motivated to financially support airlines. This is so they can maintain network connectivity to protect economic activities and jobs in the aviation industry itself, as well as in related industries such as tourism \citep{merkert2013determinants, gossling2017subsidies, njoya2018understanding}. 
During normal economic times, airlines receive financial support from the private sector, like banks and investors; however, these supporters may have grown more cautious when aviation industry's future became uncertain \citep{truxal2020state}. 
This indicates that government support might be the only option for airlines. 
However, several studies have pointed out that this governmental support may prioritize specific airlines, \add{especially FSCs \citep{albers2020european},} which leads to concerns about competition among airlines \citep{akbar2020bargain, abate2020government}.
\add{\cite{kaffash2023us} pointed out that the business model of LCCs allows them to open new routes on a trial-and-error basis during the pandemic. 
However, the network dynamics of different airlines have been rarely explored and are largely unknown.}

\add{Compared to the previous studies, the novelty of our study can be summarized as follows.
First, in contrast to most previous studies that analyzed air transport networks during the COVID-19 pandemic at the global, regional, or country levels, we analyze the networks for individual airlines.
As a result, we found that the structural changes in the networks during the pandemic vary significantly between different airlines, even within the same country (in this study, the US).
Second, we perform temporal network analysis for airline networks, which is greatly different from the previous studies that utilized traditional static network analysis. 
In the present study, we show that the temporal network analysis reveals evolution patterns that the static network analysis alone cannot capture.}

\section{Methods}
\label{sec:methods}

\add{We describe the methods in this section. 
First, we construct the daily network for all airlines as well as for each individual airline from the data (subsection \ref{sec:Data_and_construction_of_networks}). 
Next, we measure the changes in traditional network measurements (subsection \ref{sec:Measurements for network structure}) and assess robustness and resilience (subsection \ref{sec:Robustness and resilience}). 
Then, we elucidate the evolution patterns of the temporal network (subsection \ref{sec:Analysis of evolution patterns}). 
It is important to note that the identified evolution patterns cannot be captured by the traditional network measurements alone. 
Finally, we further analyze the recovery process of the networks at the individual flight level using link prediction (subsection \ref{sec:Link prediction}).}

\subsection{Data and construction of networks}
\label{sec:Data_and_construction_of_networks}

We use a dataset available from the US Bureau of Transportation Statistics \citep{BTS2020air}. The dataset contains the origin and destination airports of each flight, departure and arrival times, and operating airlines, among other information. We downloaded data from January 2019 to December 2021. 
In Table~\ref{tab:data}, we detail the 17 airlines included in the dataset.

From the data, we construct a daily snapshot of networks from January 2019 to December 2021, in which a node is an airport. We connect two nodes using an edge if there is at least one direct commercial flight between the two airports during the day.
\add{Note that we analyze unweighted networks because we are particularly interested in structural changes due
to the withdrawal of existing routes and the introduction of new routes.}
We construct both the entire network, including all the airlines, and the networks for each individual airline.

\begin{table}[t]
\begin{center}
\begin{threeparttable}
\caption{The 17 airlines contained in the US Bureau of Transportation Statistics dataset.}\label{tab:data}
\begin{tabular}{ccc}
\hline
Category                                    & Airline name       & International Air Transport Association code \\ \hline
\multirow{5}{*}{Full-Service Carrier (FSC)} & American Airlines  & AA        \\ \cline{2-3} 
                                            & Alaska Airlines    & AS        \\ \cline{2-3} 
                                            & Delta Air Lines    & DL        \\ \cline{2-3} 
                                            & Hawaiian Airlines  & HA        \\ \cline{2-3} 
                                            & United Airlines    & UA        \\ \hline
\multirow{5}{*}{Low-Cost Carrier (LCC)}     & JetBlue            & B6        \\ \cline{2-3} 
                                            & Frontier Airlines  & F9        \\ \cline{2-3} 
                                            & Allegiant Air      & G4        \\ \cline{2-3} 
                                            & Sprit Airlines     & NK        \\ \cline{2-3} 
                                            & Southwest Airlines & WN        \\ \hline
\multirow{7}{*}{Regional Carrier (RC)}      & Endeavor Air       & 9E        \\ \cline{2-3} 
                                            & Envoy Air          & MQ        \\ \cline{2-3} 
                                            & PSA Airlines       & OH        \\ \cline{2-3} 
                                            & SkyWest Airlines   & OO        \\ \cline{2-3} 
                                            & Mesa Airlines      & YV        \\ \cline{2-3} 
                                            & Republic Airways  & YX        \\ \cline{2-3} 
                                            & ExpressJet\tnote{*}       & EV        \\ \hline
\end{tabular}
\begin{tablenotes}
      \small
      \item[*] Note that ExpressJet ceased its operations between September 2020 and September 2021.
    \end{tablenotes}
\end{threeparttable}
\end{center}
\end{table}

\subsection{Measurements for network structure}
\label{sec:Measurements for network structure}
 We model an airline network as graph $G$ consisting of $N$ nodes (i.e., airports) and $M$ edges (i.e., direct flight connections). 
 In addition to the number of nodes and edges, we measure \add{three} indices: average degree, average clustering coefficient, and assortativity \citep{newman2018networks}. These \add{three} indices enable us to understand the characteristics of networks from different perspectives: each node's connectivity with other nodes (average degree) and tendency (assortativity), \add{and} the connectivity of each node's neighbors (average clustering coefficient).

The degree of node $i$, denoted by $k_i$, is the number of edges that node $i$ has. The average degree, denoted by $\langle k \rangle$, is the average value of the degree of all the nodes in the network:
\begin{equation}
\langle k \rangle = \frac{1}{N} \sum_{i=1}^{N} k_i\,.
\label{eq:ave_degree}
\end{equation}


The clustering coefficient for node $i$, denoted by $C_i$, quantifies the connectivity of the neighbors of node $i$, which is defined as follows:
\begin{equation}
C_i = \frac{2M_i}{k_i (k_i-1)}\,,
\label{eq:clustering}
\end{equation}
where $M_i$ is the number of edges that connect the neighbors of node $i$. 
Note that the neighbors of node $i$ include all nodes directly connected to it but exclude node $i$ itself. A higher value of $C_i$ means that the node has more connections among its neighbors. In a complete graph in which all pairs of nodes are directly connected, $C_i$ of all nodes equals $1$. Note that $C_i$ of nodes with $k_i = 1$ is defined to be $0$.
The average clustering coefficient, denoted by $C$, is the average value of the clustering coefficient of all the nodes in the network:
\begin{equation}
C = \frac{1}{N} \sum_{i=1}^{N} C_i\,.
\label{eq:ave_clustering}
\end{equation}
A higher value of $C$ implies that there are more triangles in the network and more detours between pairs of nodes.

Assortativity, denoted by $r$, is the Pearson correlation coefficient of degree between pairs of connected nodes:
\begin{equation}
r = \frac{\sum_{ij} (A_{ij} -k_ik_j/2M) k_ik_j }{\sum_{ij} (k_i\delta_{ij} -k_ik_j/2M) k_ik_j}\,,
\label{eq:assortativity}
\end{equation}
where $A_{ij}$ is the adjacency matrix, and $\delta_{ij}$ is the Kronecker delta. 
If nodes with similar values of degree tend to be connected to each other, the value of $r$ becomes positive (assortativity). Conversely, if nodes with a higher degree tend to be connected to nodes with a lower degree, then the value of $r$ becomes negative (disassortativity).

\subsection{Robustness and resilience}
\label{sec:Robustness and resilience}
In addition to the measurements in the previous section, we also analyze the robustness and resilience of individual airline networks. Robustness and resilience of transportation networks have been studied intensively since the 1990s to assess damage caused by natural and manmade disasters \citep[e.g.,][]{mattsson2015vulnerability, taylor2017vulnerability, sugishita2021vulnerability}. To evaluate robustness and resilience, it is necessary to quantitatively assess the degradation and recovery of network performance \citep{lordan2015robustness, zhou2019efficiency, zhou2021vulnerability, li2021impact, siozos2021study, sun2021robustness}. In this study, we measure network efficiency as a performance measurement, as proposed by \cite{latora2001efficient}.
Network efficiency, denoted by $E$, is defined by
\begin{equation}
E = \frac{2}{N(N-1)} \sum_{1\leq i<j\leq N} \frac{1}{l_{ij}}\,.
\label{eq:efficiency}
\end{equation}
In this definition, the reciprocal of the shortest path length is regarded as the “efficiency” of individual paths. 

\begin{figure}[tb]
\centering
\includegraphics[width=0.6\linewidth]{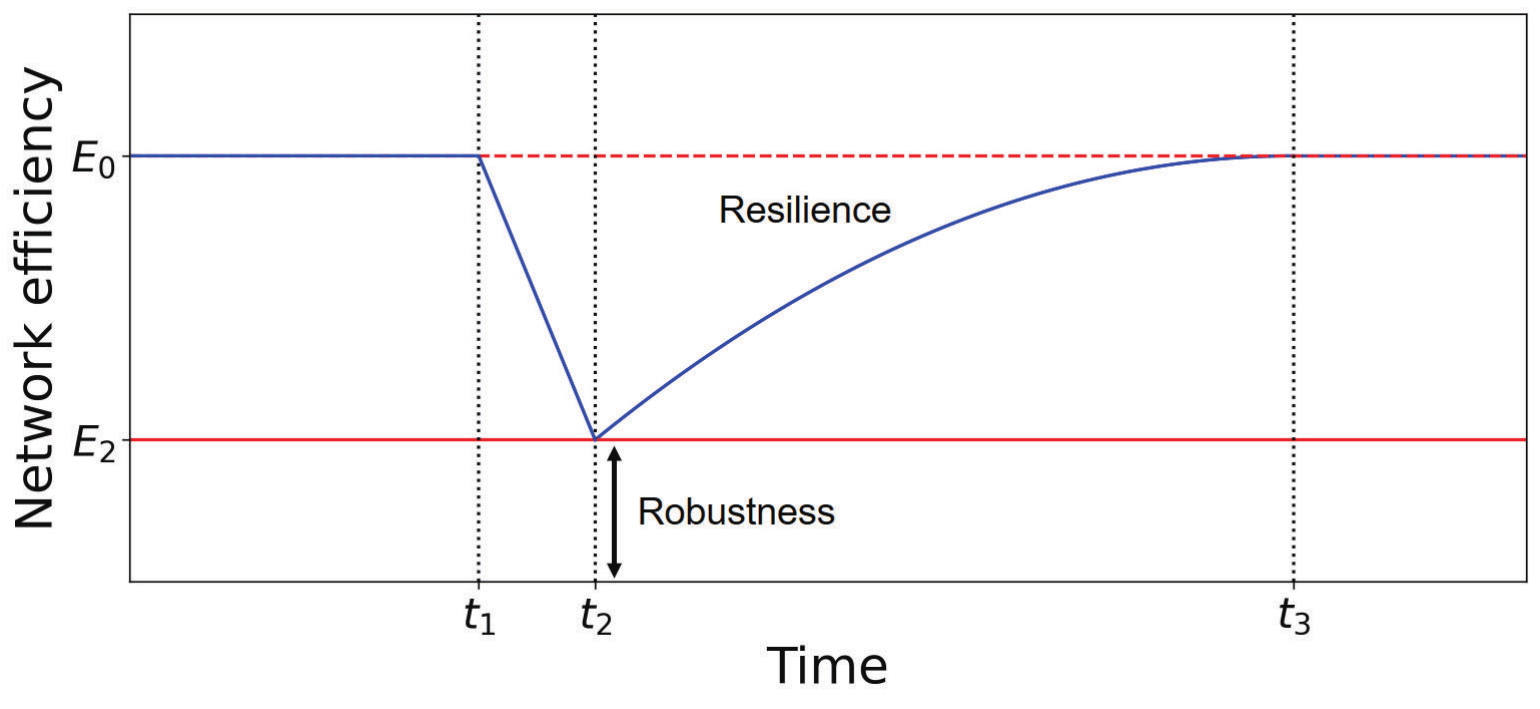}
\caption{Schematic illustration of the definition of robustness and resilience.}
\label{fig:robustness_resilience_definition}
\end{figure}

Fig.~\ref{fig:robustness_resilience_definition} shows a schematic illustration of the definition of robustness and resilience. Robustness, denoted by $R_o$, quantifies the initial impact of the COVID-19 pandemic on the loss of network efficiency, which is defined by
\begin{equation}
R_o = \frac{E_2}{E_0}\,,
\label{eq:robustness}
\end{equation}
where $E_0$ is the average value of the network efficiency in 2019, and $E_2$ is the minimum value of the seven-day moving average of network efficiency after the declaration of a national emergency. The robustness $R_o$ takes a value between 0 and 1, with closer to 1 meaning more robust.

Resilience, denoted by $R_e$, quantifies not only the decrease in network efficiency but also the recovery process \citep{bruneau2003framework}, which is defined by 
\begin{equation}
R_e = \frac{\int_{t_1}^{t_3}E(t) dt}{(t_3-t_1)E_0} \,,
\label{eq:resilience}
\end{equation}
where $t_3$ is the date when the seven-day moving average of network efficiency reaches $E_0$ for the first time since the declaration of a national emergency.
Resilience $R_e$ takes a value between 0 and 1, with closer to 1 meaning more resilient.

\subsection{Analysis of evolution patterns}
\label{sec:Analysis of evolution patterns}
Here, we apply the framework for analyzing evolution patterns in temporal networks, in which network structure changes over time \citep{masuda2019detecting, sugishita2021recurrence}, to airline networks. We measure the dissimilarity between two snapshot networks $G$ and $G^{\prime}$ using the network distance defined by
\begin{equation}
d(G,G^{\prime})= 1-\frac{M(G\cap G^{\prime})}{\sqrt{M(G)M(G^{\prime})}}\,,
\label{eq:normalized_network_distance}
\end{equation}
where $M(G)$ and $M(G^{\prime})$ are the numbers of edges in $G$ and $G^{\prime}$, respectively, and $M(G\cap G^{\prime})$ is the number of edges that $G$ and $G^{\prime}$ have in common. Network distance $d$ ranges between $0$ and $1$. The distance matrix is a $t_{\max} \times t_{\max}$ symmetric matrix of which the $(t, t')$th entry is given by $d(G_t, G_{t'})$, 
where $G_t$ is the network on day $t$, and $t_{\max}$ is the number of days observed. By analyzing the distance matrix, we can understand how each airline changes the network structure (e.g., whether the network structure returns to its pre-pandemic structure).

\subsection{Link prediction}
\label{sec:Link prediction}
To explore the recovery process of the airline networks, we perform link prediction \citep{lu2011link}. 
Link prediction is the problem of predicting the existence of an edge between two nodes in a network.
We apply two measurements to calculate scores for the link prediction, common neighbor centrality \citep{ahmad2020missing} and preferential attachment \citep{liben2003link}.
Note that a higher score means higher likelihood that the edge will appear.

Common neighbor centrality is based on two important properties of nodes, the number of common neighbors and their centrality. 
The score of the common neighbor centrality for an edge between nodes $u$ and $v$ is defined by 
\begin{equation}
\alpha (|\Gamma(u) \cap \Gamma(v)|)+(1-\alpha)\frac{N}{l_{uv}}\,,
\label{eq:common_neibor_centrality}
\end{equation}
where $\alpha$ is a parameter that varies between 0 and 1, and $\Gamma(i)$ is the set of neighbors of node $i$.
The first term represents the number of common neighbors. 
The second term represents the closeness centrality between two nodes in a network with $N$ nodes, which is defined by \citet{ahmad2020missing}.

Preferential attachment is based on the idea that 
The score of the preferential attachment for an edge between nodes $u$ and $v$ is defined by 
\begin{equation}
|\Gamma(u)||\Gamma(v)|\,.
\label{eq:preferential_attachment}
\end{equation}

\section{Results}
\label{sec:results}

\begin{figure}[b!]
\centering
\includegraphics[width=\linewidth]{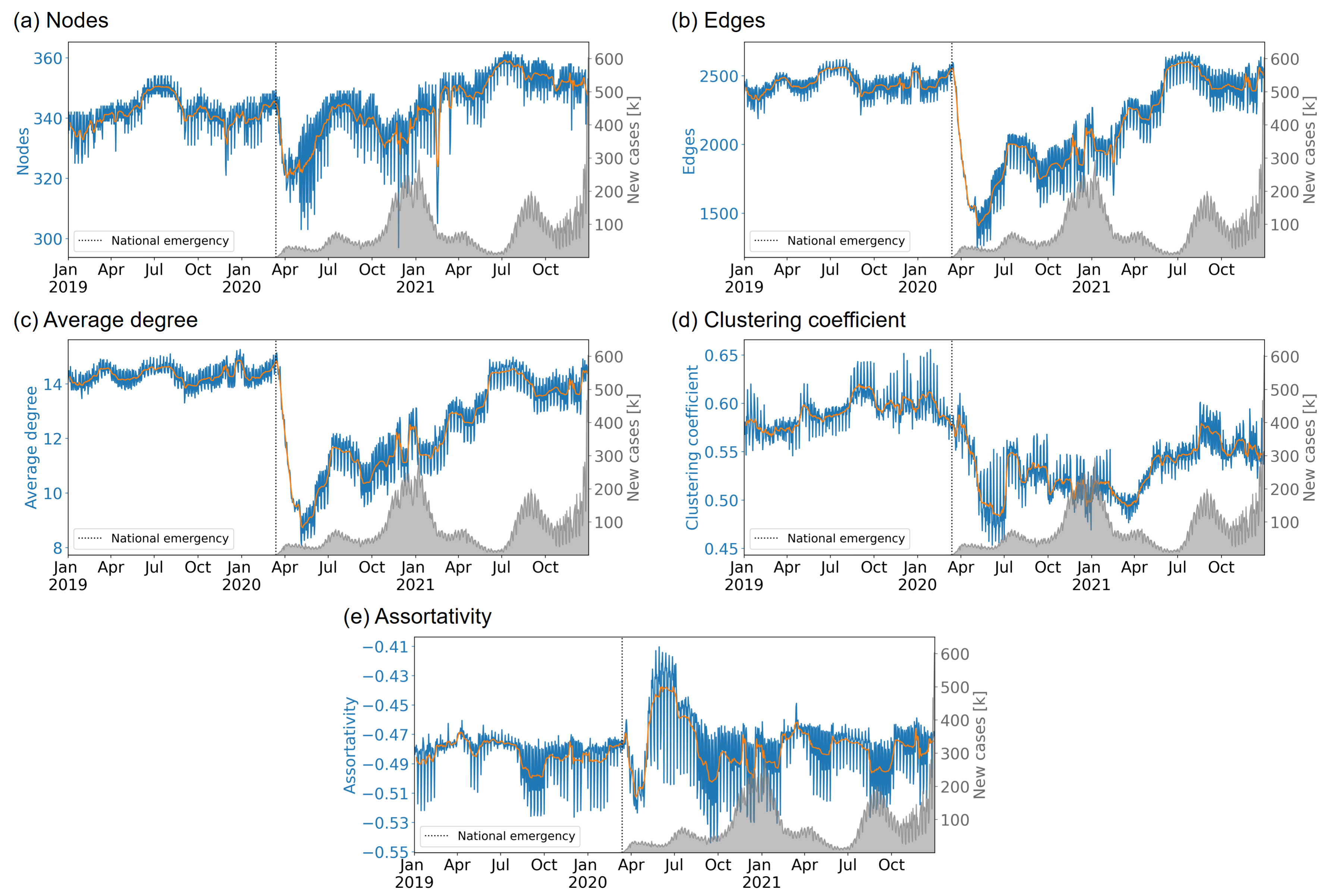}
\caption{Evolution of the measurements for \add{the entire US domestic airline network}: (a) nodes, (b) edges, (c) average degree, \add{(d)} average clustering coefficient, and \add{(e)} assortativity.
The orange curves show the seven-day moving average of each measurement.
We also show the number of new cases of COVID-19 per day in the US in gray.}
\label{fig:entire_network}
\end{figure}

\subsection{The entire US domestic airline network}
We show the changes in the \add{five} indices (i.e., nodes, edges, average degree, average clustering coefficient, and assortativity) for the entire US domestic airline network in Fig.~\ref{fig:entire_network}. We also show the number of new cases of COVID-19 per day in the US in gray \citep{CDC2021covid}. All the measurements fluctuate slightly on a daily basis. We can observe periodic patterns, which can be considered weekly periodicity. To smooth out this weekly periodicity, we show the seven-day moving average of each measurement as orange curves. The dotted line indicates the declaration of the national emergency on March 13, 2020.

All the indices significantly change after the declaration of the national emergency. The number of nodes fall by approximately 5\% and return to its pre-pandemic level around July 2020. The number of edges then drops sharply by approximately 50\% and then gradually increases, finally returning to its pre-pandemic level around July 2021. The variation in the average degree is similar to that of the number of edges. This means that it has taken more than one year for the average number of direct flight connections at each airport to return to that of the pre-pandemic level. This result also indicates that the value is continuously increasing and moving toward recovery, although it seems somewhat affected by the number of new cases (especially around a large peak of new cases in January 2021). The average clustering coefficient meanwhile, is slightly lower than the pre-pandemic level even in 2021. 
This means that there are fewer triangles in the network and fewer detours. 
Assortativity takes negative values of approximately $-0.4$ for all periods. This means that high-degree airports tended to connect with low-degree airports. There is a decrease immediately after the national emergency, followed by a large increase. This could be due to the loss of flights between the high-degree and the low-degree nodes, resulting in an increase of assortativity. It then decreases from around July 2020 and returns to its pre-pandemic level around October 2020.

\begin{figure}[t!]
\centering
\includegraphics[width=0.55\linewidth]{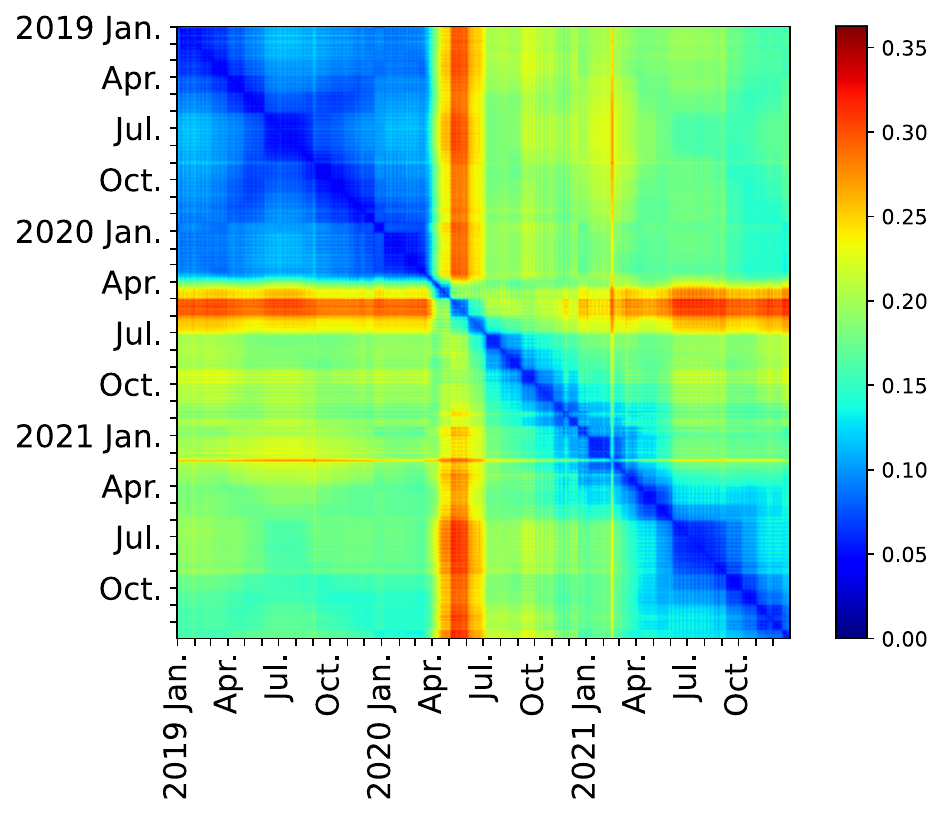}
\caption{Distance matrix for the entire US domestic airline network. The color indicates the network distance $d$ between pairs of networks.}
\label{fig:entire_network_distance_matrix}
\end{figure}

Next, we show the distance matrix for the entire US domestic airline network in Fig.~\ref{fig:entire_network_distance_matrix}. The colors indicate the values of network distance $d$ defined in Eq.~\eqref{eq:normalized_network_distance}. Higher values indicate more dissimilar networks. In Fig.~\ref{fig:entire_network_distance_matrix}, we can observe a blue block in the upper left corner of the distance matrix. This indicates that before the national emergency was declared, the network is relatively similar, although there are monthly or seasonal variations. However, around the declaration of the national emergency, we can see red rectangles. This suggests a dramatic change in the network, which is consistent with the significant changes in the network measurements in Fig.~\ref{fig:entire_network}.
Furthermore, most interestingly, we can observe that the structure of the network during the pandemic was largely different from the pre-pandemic network structure. In other words, even though the number of nodes and edges had already returned to their pre-pandemic levels in July 2021 (see Fig.~\ref{fig:entire_network}a-b), the network structure was still largely different from that from before the pandemic. 
In the next sections, we examine the dynamics of the individual airline networks.

\subsection{Networks for 16 airlines}
In this section, we analyze changes in the network structure of each of the 16 airlines in the dataset (see also Table~\ref{tab:data}). Note that we excluded ExpressJet from the analyses in this section because it ceased its operations from September 2020 to September 2021.

\subsubsection{Evolution patterns}
\begin{figure}[ht!]
\centering
\includegraphics[width=\linewidth]{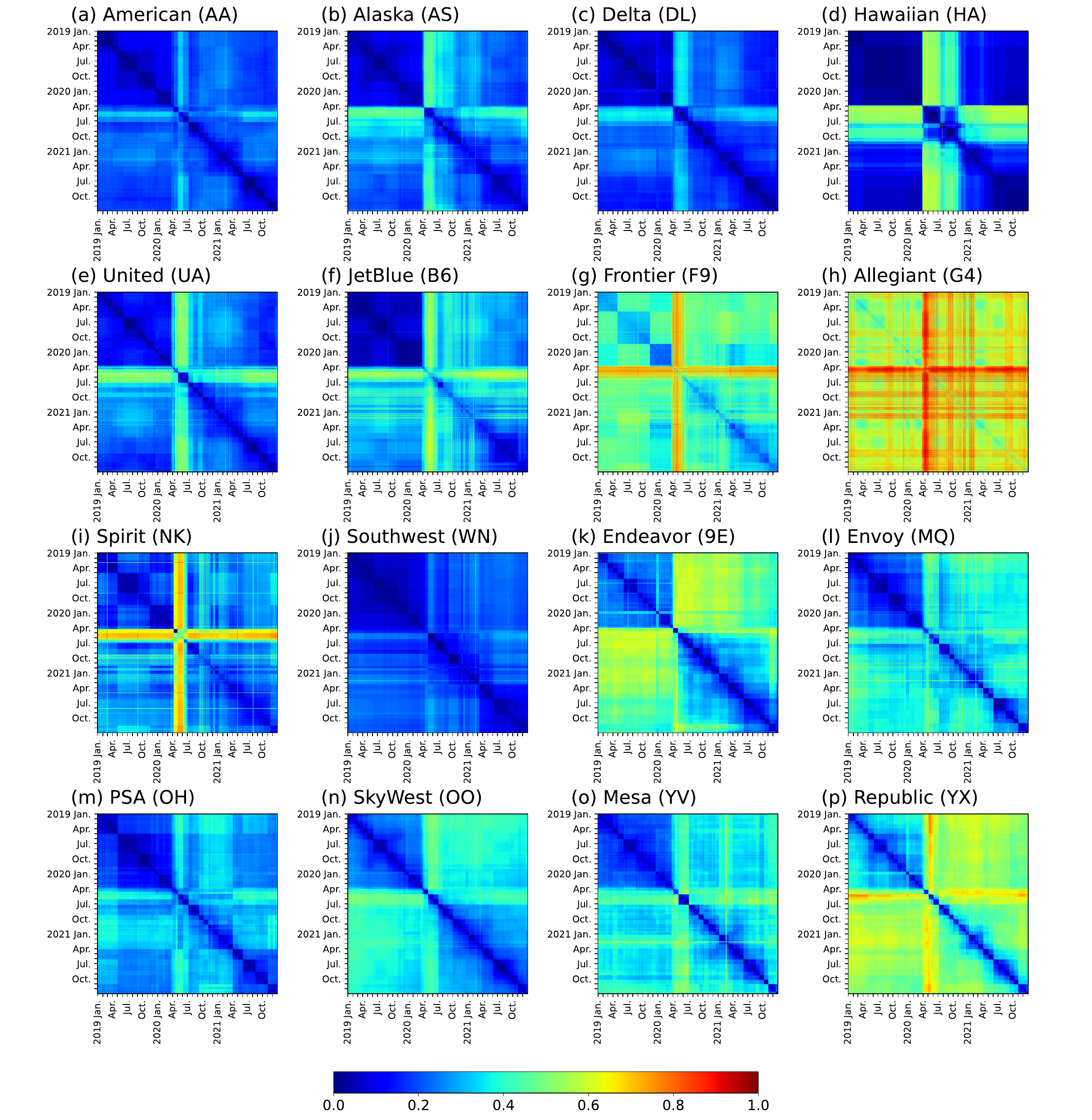}
\caption{Distance matrices for 16 airlines. The color indicates network distance $d$ between pairs of networks.}
\label{fig:Distance_matrix16}
\end{figure}

We show the distance matrices for the 16 airlines in Fig.~\ref{fig:Distance_matrix16}. Recall that five airlines are FSCs (i.e., American, Alaska, Delta, Hawaiian, and United), five airlines are LCCs (i.e., JetBlue, Frontier, Allegiant, Spirit, and Southwest), and six airlines are RCs (i.e., Endeavor, Envoy, PSA, SkyWest, Mesa, and Republic) (see also Table~\ref{tab:data}). The results indicate that there are similarities and differences in changes in the network structure of each airline.

First, for the five FSCs (Figs.~\ref{fig:Distance_matrix16}a-e), we observe that the network structure tends to return to the pre-pandemic structure a few months after the national emergency was declared. This is common to all five FSCs, although the timing of their return differs (e.g., Hawaiian took approximately seven months to return, as shown in Fig.~\ref{fig:Distance_matrix16}d). There are several possible reasons for this relatively early return to the pre-pandemic structure. 
One is empty planes being flown to avoid losing slots at slot-constrained airports, `ghost flights’ \citep{sun2022ghostbusters}. In fact, Haanappel (2020) pointed out that some airlines did this. There could also have been sole transportation of cargo. Furthermore, it is likely that they received financial aid from the government to maintain network connectivity. In fact, each of the three major FSCs (i.e., American, Delta, and United) received more than \$5 billion from the US Treasury Department between January and August 2020 \citep{statista2022government}. \cite{neal2011business} showed that FSCs have less flexible business models and are more affected by uncertain events than LCCs. Our results further suggest that FSCs may also be inflexible in terms of network structure during emergent situations.

\begin{figure}[t!]
\centering
\includegraphics[width=\linewidth]{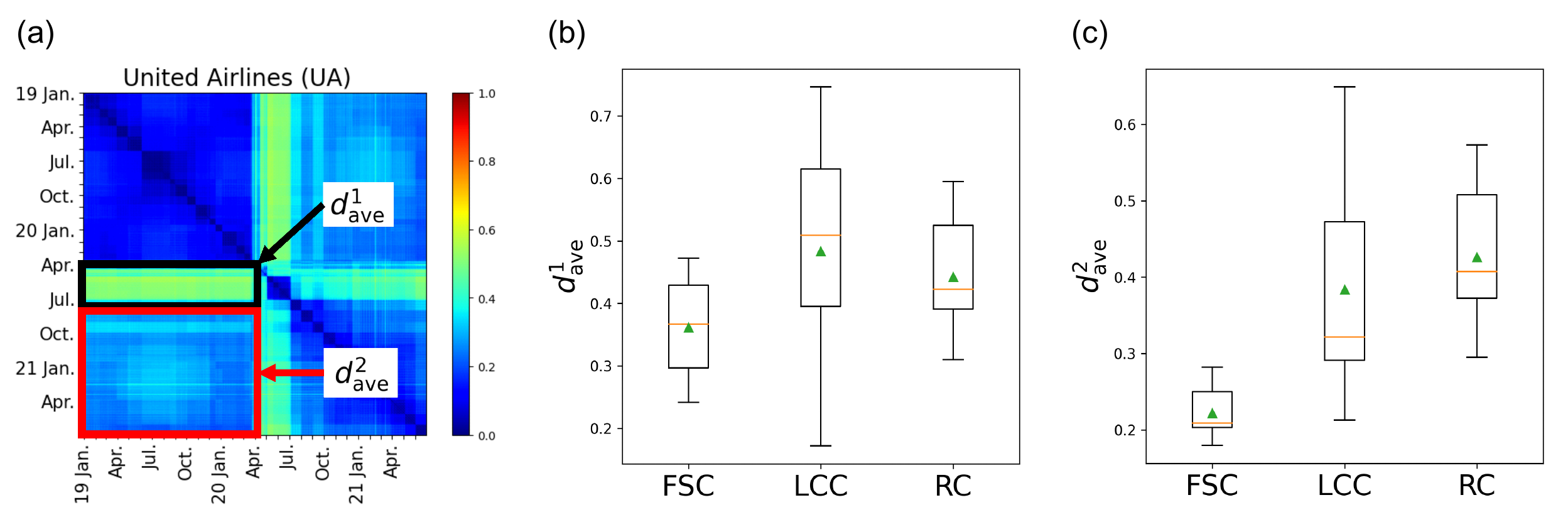}
\caption{Quantitative comparison of changes in network structure among full-service carriers (FSCs), low-cost carriers (LCCs), and regional carriers (RCs). (a) The schematic illustration of the definition of $d_{\rm ave}^1$ and $d_{\rm ave}^2$. 
(b) Comparison of $d_{\rm ave}^1$ among FSCs, LCCs, and RCs. We use box plots to show summaries of five distribution values: the first quartile ($Q_1$), the median, the third quartile ($Q_3$), the minimum ($Q_1-1.5 \, \times \, {\rm IQR}$), and the maximum ($Q_3+1.5 \, \times \, {\rm IQR}$), where ${\rm IQR}=Q_3-Q_1$. 
(c) Comparison of $d_{\rm ave}^2$ among FSCs, LCCs, and RCs.}
\label{fig:analysis_distance_matrix}
\end{figure}

Second, for the five LCCs (Figs.~\ref{fig:Distance_matrix16}f-j), we observe significant changes, represented by the vivid red lines immediately after the declaration of a national emergency, except for Southwest. This is especially clear for Frontier, Allegiant, and Spirit (Figs.~\ref{fig:Distance_matrix16}g-i). 
This result suggests that LCCs changed their network structure more dramatically than FSCs or RCs immediately after the national emergency was declared. We then compare such changes by airline type. As shown in Fig.~\ref{fig:analysis_distance_matrix}a, we define $d_{\rm ave}^{1}$ as the average over the values of the network distance $d$ between March 13 and June 30, 2020, which is shown by a black rectangle in Fig.~\ref{fig:analysis_distance_matrix}a. In Fig.~\ref{fig:analysis_distance_matrix}b, we show the results of the comparison of $d_{\rm ave}^{1}$ among FSCs, LCCs, and RCs. The results show that LCCs tend to have larger values of $d_{\rm ave}^{1}$ than FSCs and RCs.
In addition, we also notice that each airline has other unique characteristics. Allegiant has no clear blue blocks, which suggests that it had already been frequently changing its network structure before the pandemic. This result is consistent with the large fluctuations in the measurements for network structure shown in \ref{sec:each_airline_measurements} (e.g., Fig.~\ref{fig:Nodes_16}h). Furthermore, Southwest had been evolving in a very unique way. That is, there is relatively little fluctuation immediately after the national emergency. In fact, even during the COVID-19 pandemic, among the 16 airlines, only Southwest expanded its network by increasing the number of nodes (see Fig.~\ref{fig:Nodes_16}j). Previous studies have also noted Southwest's distinctive characteristics \citep[e.g.,][]{czaplewski2001southwest, smith2004evaluation,  sugishita2021recurrence}, and our study quantitatively observes one more unique feature that arose during the COVID-19 pandemic.

Finally, for the six RCs (Figs.~\ref{fig:Distance_matrix16}k-p), we observe that these airlines significantly changed their network structure and tended to shift to the new structure during the pandemic. Even though PSA returned to some extent to its pre-pandemic structure around April 2021, all the other RCs shifted to a new structure (especially Endeavor, SkyWest, and Republic). In the same way as $d_{\rm ave}^{1}$, we define $d_{\rm ave}^{2}$ as the average value of network distance $d$ between July 2020 and December 2021, which is shown by a red rectangle in Fig.~\ref{fig:analysis_distance_matrix}a. In Fig.~\ref{fig:analysis_distance_matrix}c, we show the results of the comparison of $d_{\rm ave}^{2}$ among FSCs, LCCs, and RCs. The quantitative results are consistent with our observations, which suggest that RCs tend to have larger values of $d_{\rm ave}^{2}$ than FSCs and LCCs. In general, RCs are subsidiaries of the FSCs (e.g., Endeavor operates as Delta Connection for Delta). Our results suggest that FSCs, which cannot flexibly change their network structure, have a strategy to significantly change the network structure of corresponding RCs.

\subsubsection{Robustness, resilience, and financial loss}
In this section, we evaluate the robustness and resilience of each airline and analyze its relationship with revenue loss. The revenue loss for each airline is defined as the difference between operating revenue between April 2019 and March 2020 and that between April 2020 and March 2021. We used data from the Bureau of Transportation Statistics \citep{BTS2020air}.

\begin{figure}[t!]
\centering
\includegraphics[width=0.9\linewidth]{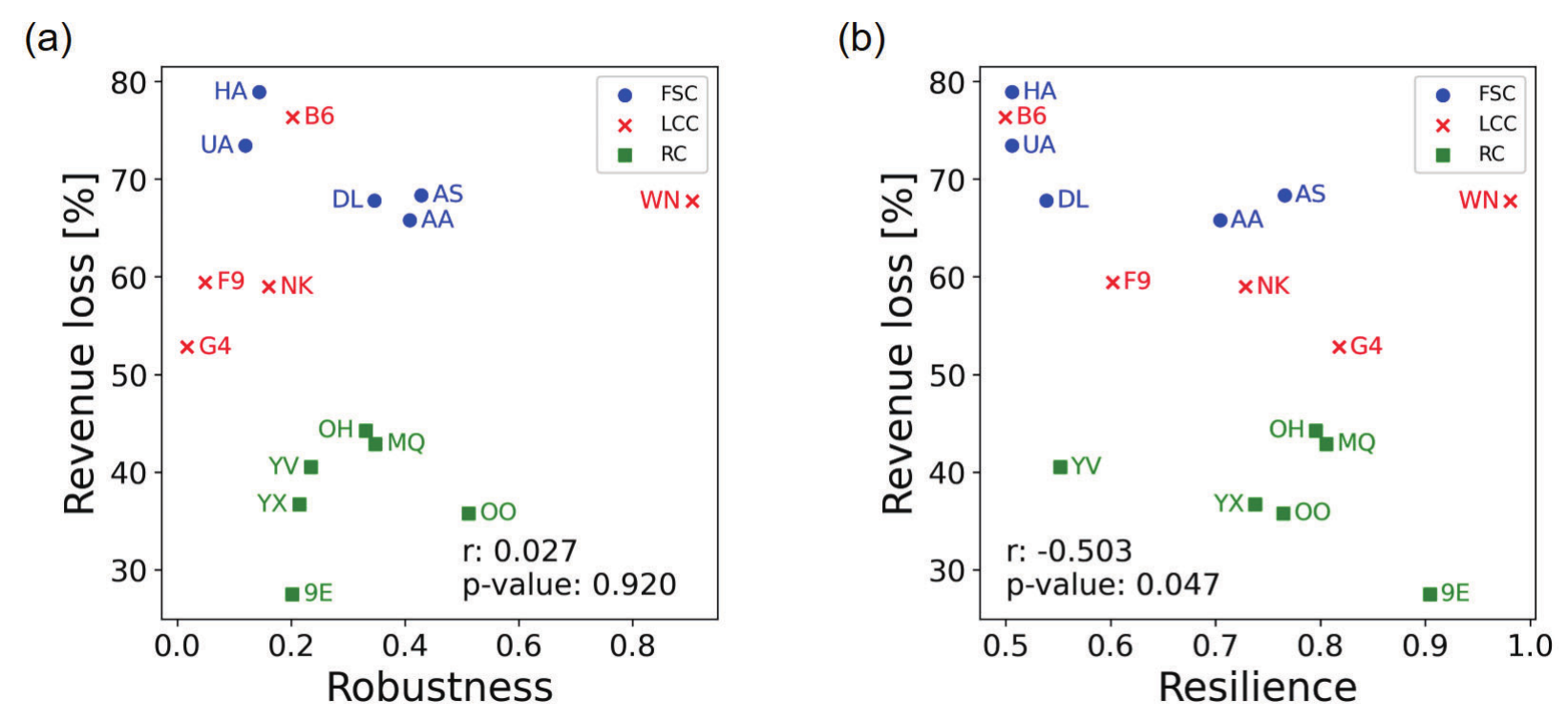}
\caption{Relationships between robustness, resilience, and revenue loss.  (a) The robustness and revenue loss. (b) The resilience and revenue loss. $r$ represents the correlation coefficient.}
\label{fig:robustness_resilience_financial_loss}
\end{figure}

Figure~\ref{fig:robustness_resilience_financial_loss} shows the results of the correlation analysis between robustness, resilience, and revenue loss. First, we see that the revenue loss itself varies widely across airline types. The revenue loss for FSCs is approximately 70\%, which is larger than that for LCCs and RCs. Hawaiian (HA), which took longer to return to its pre-pandemic structure among the FSCs, had the largest revenue loss of approximately 80\%. 
Following the FSCs are the LCCs, with losses of approximately 60\%. We also note that JetBlue (B6) and Southwest (WN) have revenue losses as high as those of the FSCs. On the other hand, RCs' revenue loss is approximately 30-40\%, which is clearly smaller than that of FSCs and LCCs. These differences regarding revenue loss among FSCs, LCCs, and RCs were also observed after the 9/11 terrorist attacks \citep{franke2004competition}. While further studies are desirable, this loss may be a universal occurrence during emergent situations. 
Second, we find no correlation between robustness and revenue loss (Fig.~\ref{fig:robustness_resilience_financial_loss}a). Additionally, while there is no clear difference in robustness by airline type, Southwest shows high robustness: even though Southwest has a relatively high revenue loss of about 70\%, the initial impact of the pandemic on network efficiency was limited. On the other hand, we find a negative correlation between resilience and revenue loss (Fig.~\ref{fig:robustness_resilience_financial_loss}b). In other words, the faster network efficiency recovers, the smaller the revenue loss, which is reasonable. However, here again, Southwest is unique. Specifically, Southwest quickly recovers its network structure despite the large revenue loss as noted earlier. As we already pointed out, Southwest is the only one of the 16 airlines that expanded its network by increasing the number of nodes, but this may be a network expansion in anticipation of future revenue. In summary, considering the previous section, our results suggest that FSCs are less flexible in changing their network structure and suffer higher revenue losses, while RCs may be able to reduce their revenue losses by significantly changing their network structure. LCCs fall between FSCs and RCs and are characterized by more pronounced differences between individual airlines.

\subsubsection{Link prediction}
\begin{figure}[t!]
\centering
\includegraphics[width=0.8\linewidth]{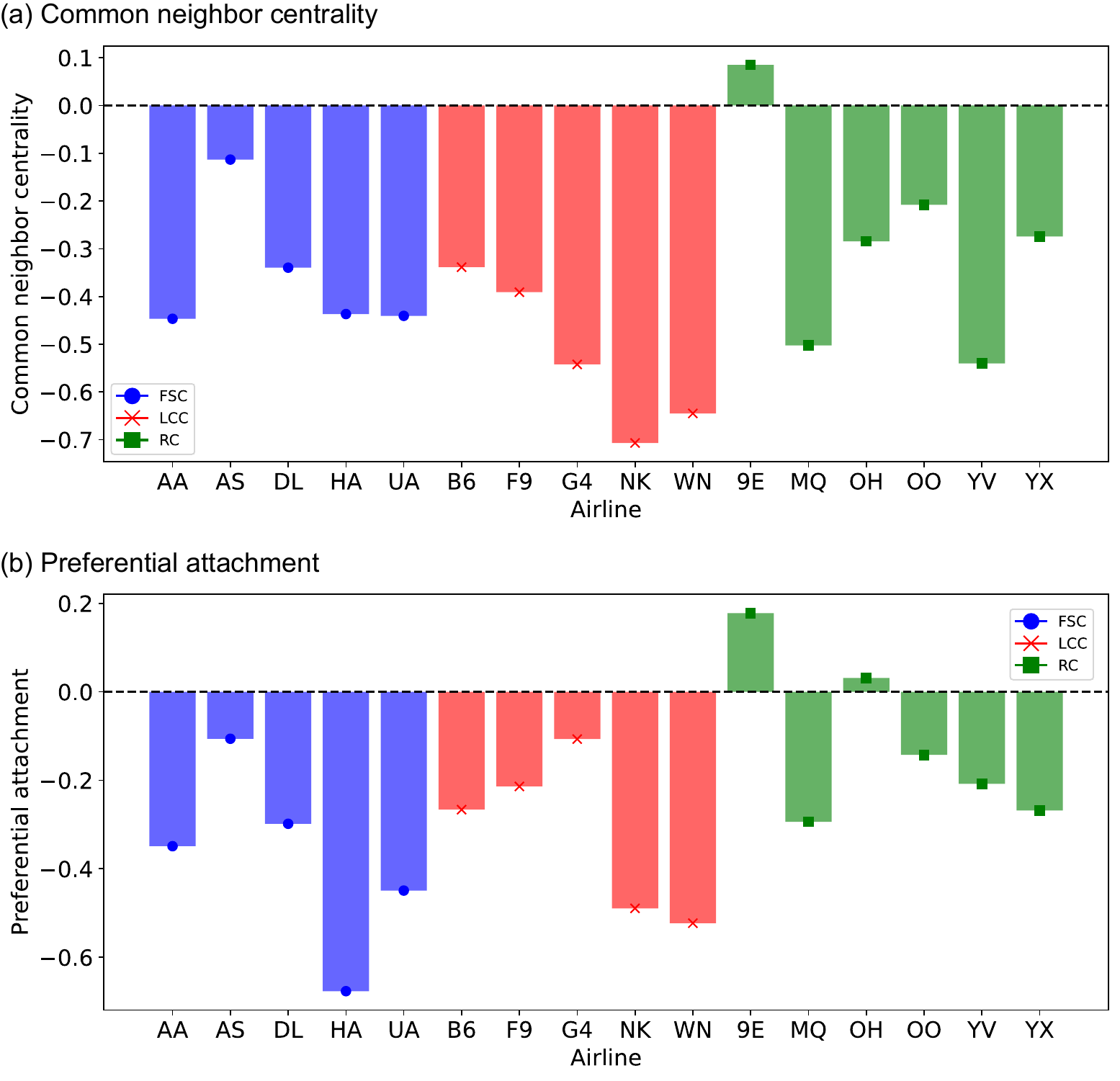}
\caption{\add{Correlation coefficients between the link prediction scores and the actual time taken for edges to recover for two measurements of the link prediction, (a) common neighbor centrality and (b) preferential attachment}. 
}
\label{fig:link_prediction_financial_loss}
\end{figure}

In this section, we further explore the recovery process for individual airlines.
We show the results for link prediction in Fig.~\ref{fig:link_prediction_financial_loss}.
The values represent the correlation coefficient between the link prediction scores and the actual time taken for edges to recover for two measurements, common neighbor centrality and preferential attachment.
Each of these indicators is based on the following hypotheses.
First, the hypothesis for the common neighbor centrality is as follows: if two airports are more central and share more common adjacent airports in the network, the direct flight between them may recover more quickly because demand between them may also be higher. 
Second, the hypothesis for the preferential attachment is as follows: if two airports have large number of connections (i.e., hubs), the direct flight between them may recover more quickly because demand between them may also be higher.
Note that we perform the link prediction to the network of the first week of April 2019 to examine the predictability of the recovery of edges based on the network under a normal condition before the pandemic. 
For the common neighbor centrality, we set $\alpha=0.8$ because \citet{ahmad2020missing} found better performance with this value.

As we show in Fig.~\ref{fig:link_prediction_financial_loss}, the values of the common neighbor centrality are negative for all airlines except Endeavor (9E). 
The two LCCs, Sprit (NK) and Southwest (WN) have especially strong negative correlations.
Similarly, the values of the preferential attachment are negative for all airlines except Endeavor and PSA (OH).
Hawaiian (HA) has the largest negative correlation.
Overall, these results indicate that the flights connecting airports that are more central and share more common neighbors, and those connecting airports with larger degrees tend to recover earlier for most of the airlines.
However, at the same time, the results also indicate that Endeavor has a unique recovery strategy.
Recalling that Endeavor has the smallest revenue loss among all airlines (see Fig.~\ref{fig:robustness_resilience_financial_loss}), the recovery strategy may have influences on the revenue loss and further studies are required in the future.

\section{Conclusion \add{and discussion}}
\label{sec:conclusions}
In this study, we revealed the day-to-day dynamics of the US domestic airline network from January 2019 to December 2021.
\add{For} the entire US domestic airline network, we could observe the network recovery process after a sudden change following the declaration of a national emergency on March 13, 2020. The number of nodes returned to its pre-pandemic level within a few months, while the number of edges finally returned around July 2021. However, we found that, even though the number of nodes and edges returned to pre-pandemic levels by July 2021, the network structure was still largely different from what it was pre-pandemic. 
\add{Note that the identified structural changes cannot be captured by the traditional static network analysis alone.}
By analyzing the network changes for each of the 16 airlines, we further determined that the network dynamics varied greatly from airline to airline. 
\add{Previous studies showed that the recovery speed of air transport varies depending on the countries \citep{gudmundsson2021forecasting,li2021spatiotemporal, sun2023data}, but our results further suggested that the network dynamics are greatly different depending on the airlines even in one country.}
Specifically, we found that all the FSCs, which suffered severe revenue loss, returned to their pre-pandemic structure within a few months after the national emergency was declared. In contrast, most of the RCs shifted to completely new networks during the pandemic, which may have contributed to reduced revenue losses. 
Furthermore, we also found that many of the LCCs made drastic changes in their network structure immediately after the national emergency was declared. 
At the same time, LCCs were characterized by more pronounced differences between airlines: Southwest in particular was the only one of the 16 airlines to expand its network despite its large revenue losses.
Finally, we further explored the recovery process based on the link prediction. We found that the flights connecting airports that are more central and share more common neighbors, and those connecting airports with larger degrees tend to recover earlier for most of the airlines. 
We also found that Endeavor, which had the smallest revenue loss, had a unique recovery strategy. 

\add{
In the previous studies, discussions on the performance during the pandemic have taken place, especially in the context of FSCs vs LCCs.
While the benefits of governmental support for FSCs are also pointed out \citep{abate2020government}, several studies argued that LCCs tend to have better financial performance during times of uncertainty due to their flexibility \citep{fontanet2022impact, sun2022startups, kaffash2023us}.
In the present study, we have quantitatively captured the differences in flexibility from a network structure perspective. 
Moreover, the results further suggested that relying solely on the category of FSC and LCC is not sufficient to discuss policies for managing air transport system during the pandemic. 
For example, the changes in network structure for Hawaiian Airlines are entirely different from the other FSCs.
Similarly to FSCs, network dynamics vary significantly among LCCs. 
\cite{kaffash2023us} argued that the business model of LCCs allows them to open new routes on a trial-and-error basis during the pandemic.
Even though our results support this argument, the results also highlight the unique strategic decisions of each LCC. 
For example, Southwest Airlines has expanded its network by adding new routes to its existing network.
It seems that Southwest is particularly well positioned to take advantage of this period of uncertainty. 
Further research is required to understand the long-term benefits of Southwest's network expansion identified in this study.
Note that network expansion has also been observed for big three European LCCs, Ryanair, easyJet, and Wizz Air \citep{zhang2023big}.
On the other hand, Frontier Airlines has shifted network structure, even though the number of nodes and edges is almost the same as before the pandemic.
Furthermore, our study has demonstrated that RCs exhibit even greater flexibility than LCCs. While RCs were predicted to be short-term winners in terms of financial performance during the early stages of the pandemic \citep{suau2020early}, we have captured distinctive route development patterns from a network structure perspective. 
However, it is worth noting that even within the RC category, there are significant variations in how networks recover.
In conclusion, although discussions have historically revolved around the categories of FSCs, LCCs, and RCs, there is a need for more individualized discussions to prepare for and respond to the pandemic.}

\add{Our results also provide valuable insights for simulating the spread of infectious diseases. 
Our results showed that the structure of the air transport network changes greatly over time and is completely different from the pre-pandemic structure, mainly due to LCCs and RCs. 
This suggests that when simulating the spread of infectious diseases to establish policies in response to the pandemic, it is crucial to grasp the current and future network structure as input. 
Failure to do so may lead to entirely different outcomes. 
For example, simulating based on the pre-pandemic network to simplify the analysis could yield meaningless results.
On the other hand, once the network has experienced a breakdown, there may be some predictability regarding the sequence in which edges are restored.
Applying the insights obtained by the link prediction in this study and simulating disease spread within the predicted network is an interesting future direction. 
By understanding how the network structure evolves during and after a crisis like a pandemic, policymakers and researchers can better prepare and respond to infectious disease outbreaks with more accurate models and strategies.}

There are many \add{other} interesting future directions that build on
our results. 
First, in our study, we have not been able to analyze how networks for individual airlines influence each other.
Analyzing competition between airlines from network changes in emergency situations could be an interesting line for future research.
For example, our study found that RCs have significantly changed their network structure, but whether this change has led to a more similar network to other airlines' networks—increasing competition on certain routes, or whether it has led to a more dissimilar network to others, opening up completely new routes—is currently unclear.
We can reveal this by analyzing distance matrices constructed by comparing networks across airlines over time \citep[see,][]{sugishita2021recurrence}.
This is also relevant to the strategy of Southwest, which was the only airline to expand its network.
\add{Second,} we have not been able to determine how the network changed geographically. 
Integrating geographical factors into the temporal network analysis warrants future work.
\add{Third, we have analyzed unweighted networks because we are especially interested in the structural changes due to the withdrawal of existing routes and the introduction of new routes. 
Analyzing the disruption and recovery process in consideration of total traffic volume and load factor is an interesting future direction. 
Changes in the aircraft mix are also worth investigating.}
\add{Fourth, it seems that the return of FSCs to their pre-pandemic network structure is influenced by subsidies provided by the federal government. While challenging, it is an important task to assess the network structural changes that would occur if subsidies were eliminated.}
\add{Fifth,} an analysis of airlines in countries other than the US during the COVID-19 pandemic should be conducted, as well as network analyses for other emergent situations such as other disease outbreaks, financial crises, terror-related events, and natural disasters. We found that the network dynamics of FSCs, LCCs, and RCs differed significantly during the COVID-19 pandemic, but this might be a universal occurrence regardless of country or type of emergency, which further research could help resolve.
We have also revealed how the airlines have recovered their networks. As a future direction, optimal recovery strategies are worth investigating for the next pandemic to come.

\appendix 
\section{Measurements of network structure for individual airlines}
\label{sec:each_airline_measurements}
\begin{figure}[H]
\centering
\includegraphics[width=\linewidth]{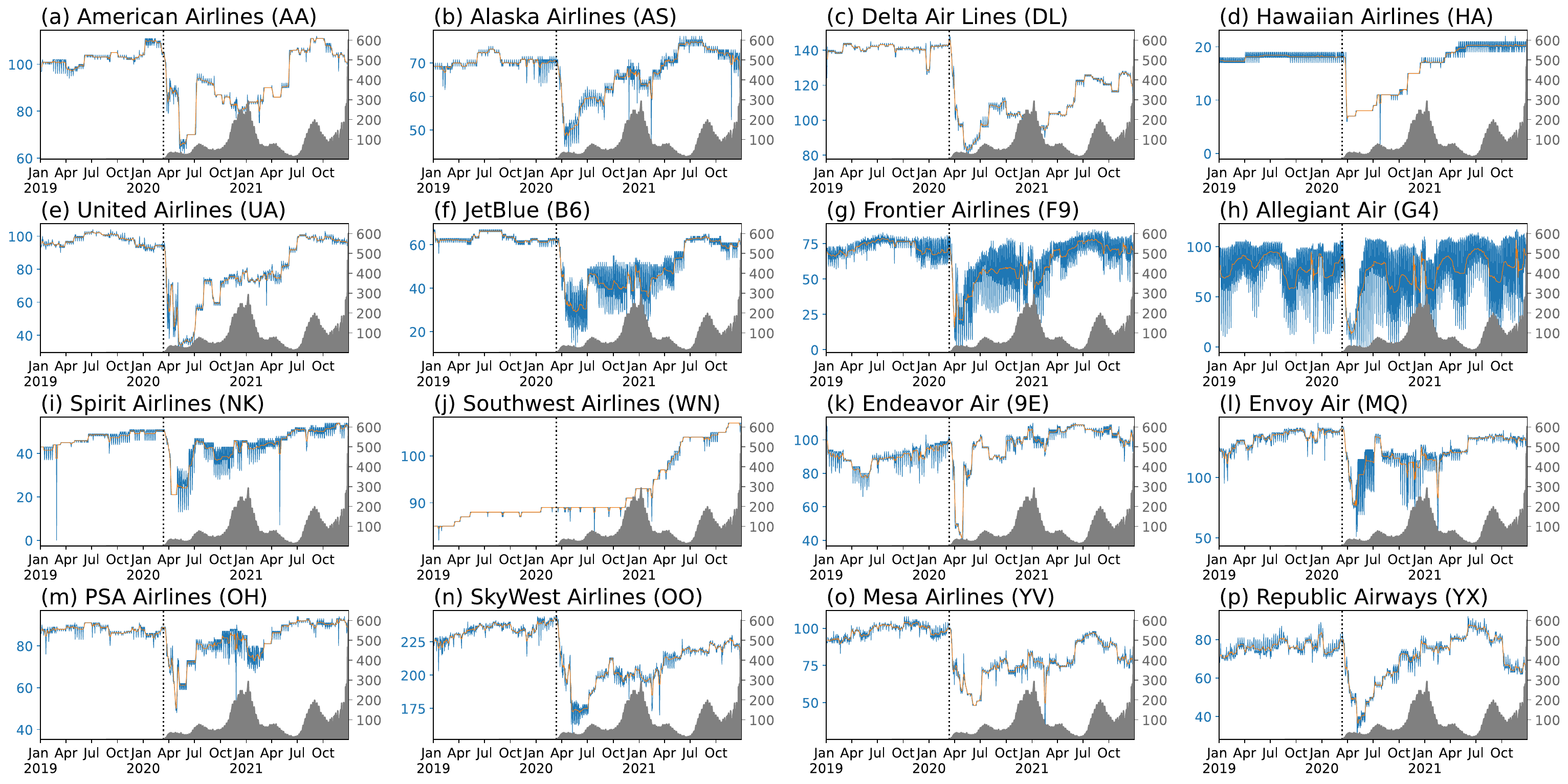}
\caption{Evolution of the number of nodes for 16 airlines.}
\label{fig:Nodes_16}
\end{figure}

\begin{figure}[H]
\centering
\includegraphics[width=\linewidth]{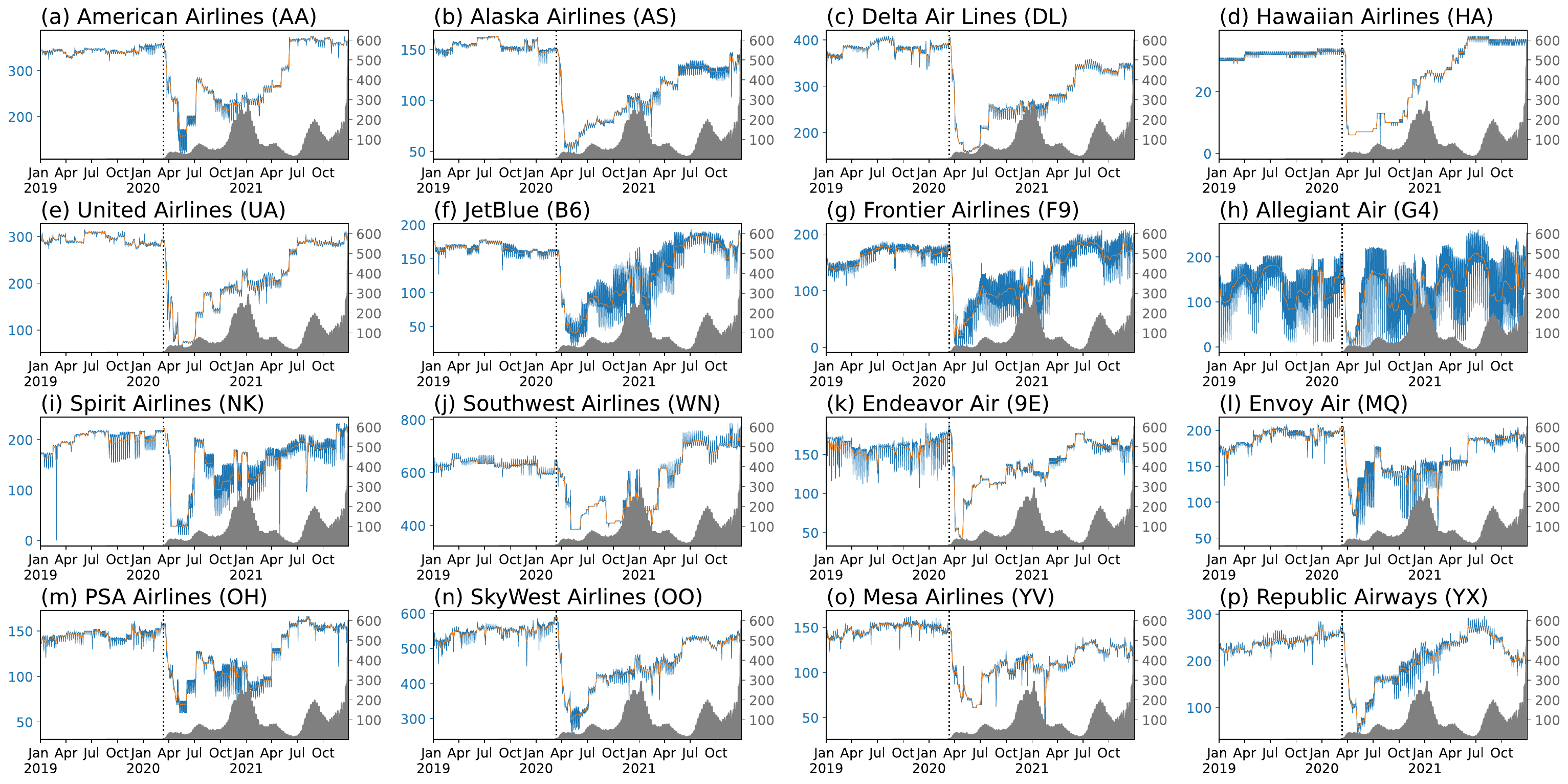}
\caption{Evolution of the number of edges for 16 airlines.}
\label{fig:Edges_16}
\end{figure}

\begin{figure}[H]
\centering
\includegraphics[width=\linewidth]{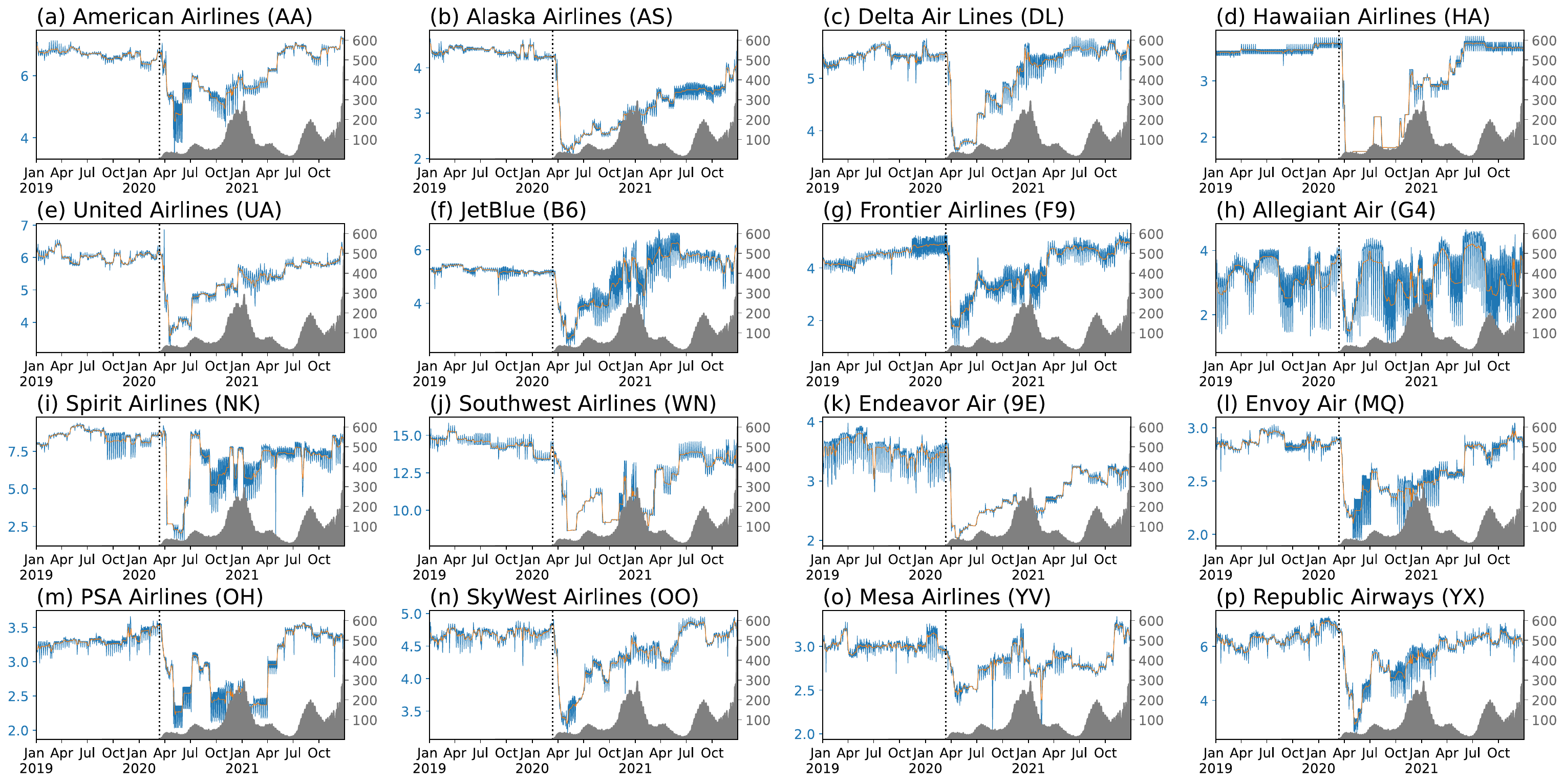}
\caption{Evolution of the average degree for 16 airlines.}
\label{fig:AD_16}
\end{figure}

\begin{figure}[H]
\centering
\includegraphics[width=\linewidth]{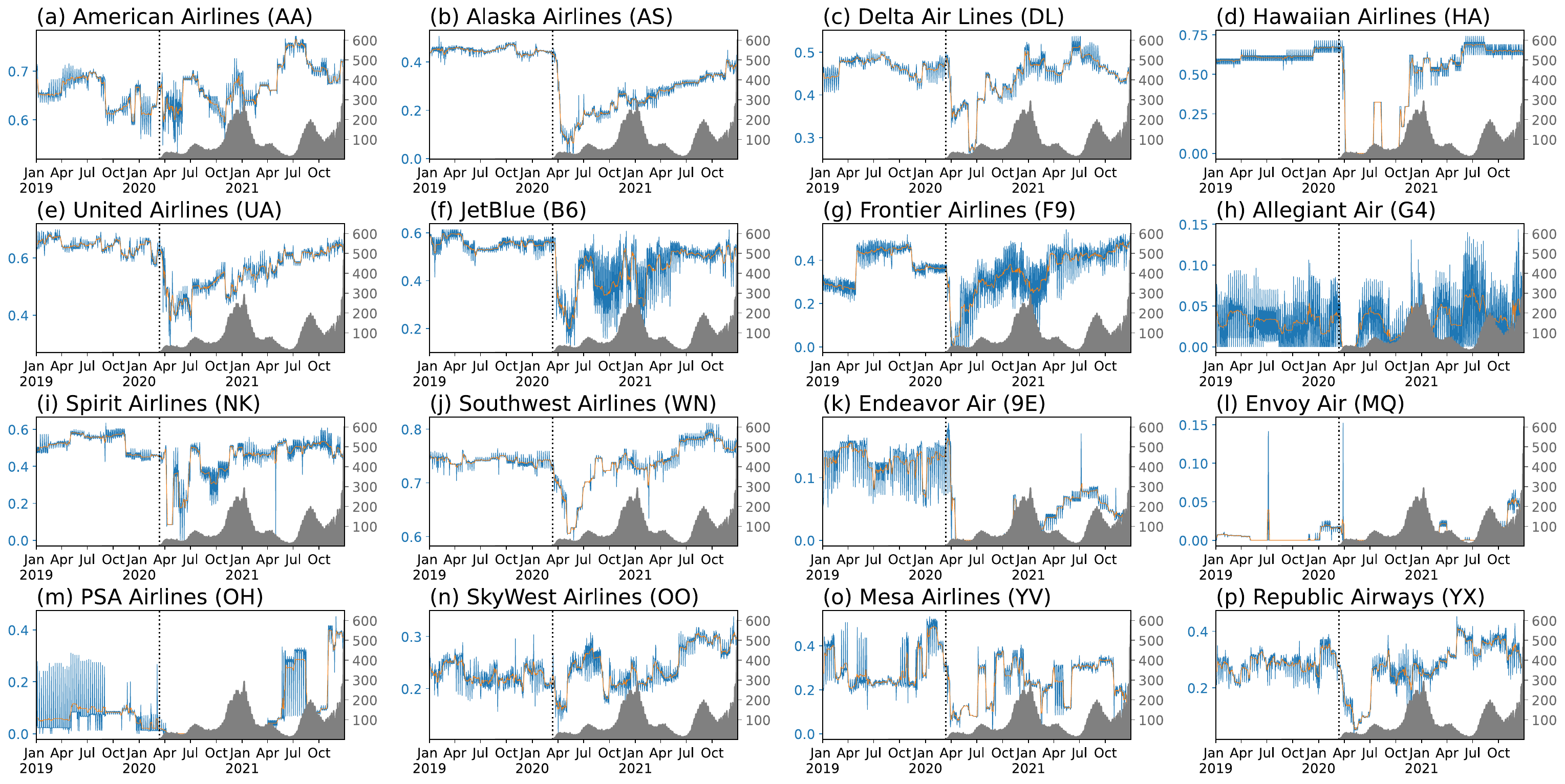}
\caption{Evolution of the average clustering coefficient for 16 airlines.}
\label{fig:CLU_16}
\end{figure}

\begin{figure}[H]
\centering
\includegraphics[width=\linewidth]{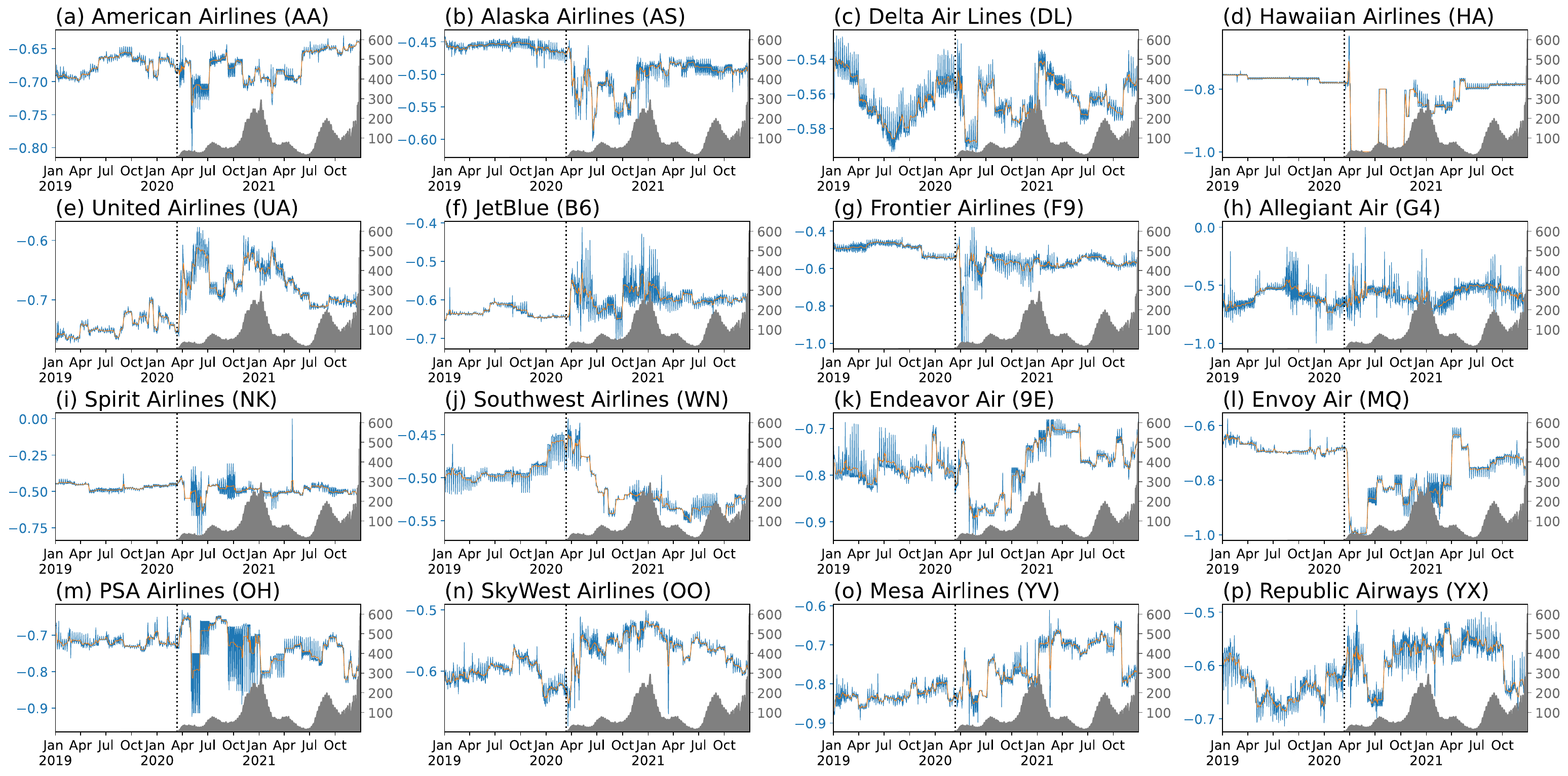}
\caption{Evolution of the assortativity for 16 airlines.}
\label{fig:ASS_16}
\end{figure}

In this section, we show the evolution of the measurements for the network structure of individual airlines (Figs~\ref{fig:Nodes_16}-\ref{fig:ASS_16}). In the same way as in Fig.~\ref{fig:entire_network}, the gray curves represent the number of new COVID-19 cases per day in the US, and the orange curves represent the seven-day moving average of each measurement. The dotted line indicates the declaration of a national emergency on March 13, 2020.

We summarize some notable characteristics for individual airlines. First, as we already pointed out, Southwest has a unique evolution pattern. That is, the number of nodes does not change after the declaration of a national emergency and then increases (Fig.~\ref{fig:Nodes_16}j). The number of edges decreases slightly but then increases (Fig.~\ref{fig:Edges_16}j). In other words, among the 16 airlines, only Southwest expanded its network during the pandemic. 
Next, we can see that Allegiant had already been frequently changing its network structure before the pandemic. All indicators are highly variable over the entire period. 
In addition, Endeavor recovers its number of nodes early and number of edges around July 2021 (Fig.~\ref{fig:Nodes_16}k and Fig.~\ref{fig:Edges_16}k), but the clustering coefficient drops significantly compared to the pre-pandemic period (Fig.~\ref{fig:CLU_16}k), suggesting that the network changes to one with many direct connections with few detours. 
Furthermore, we see that some airlines increase their weekly fluctuations after the pandemic. For example, when we examine the number of edges, they are particularly pronounced for LCCs and RCs (Fig.~\ref{fig:Edges_16}). This suggests that these airlines became more sensitive to daily demand fluctuations and changed their network accordingly. 

\section*{Acknowledgements}
This work was supported by JSPS KAKENHI Grant Number 19K23531.

\bibliographystyle{elsarticle-harv}\bibliography{ref}
\end{document}